\documentclass{PoS}

\usepackage[utf8]{inputenc}
\usepackage{amsmath,amsfonts,amssymb,mathrsfs,fixmath}
\usepackage[sort&compress]{natbib}
\bibpunct{[}{]}{,}{n}{}{}
\newcommand{\lr}[1]{ \left( #1 \right) }

\newcommand{\lrc}[1]{ \left\{ #1 \right\} }
\newcommand{\vev}[1]{ \langle \, #1 \, \rangle }

\newcommand{\mua}{\ensuremath{\mu_5}}        
\newcommand{\scme}{\ensuremath{\sigma_{CME}}} 

\newcommand{\nm}{\ensuremath{\operatorname{nm}}}

\graphicspath{{./plots/}}

\title{Chiral Magnetic Effect in finite-size samples of parity-breaking Weyl semimetals}

\ShortTitle{CME in finite-size samples}

\author{S.~N.~Valgushev\textsuperscript{*} \\
        Institut f{\"u}r Theoretische Physik, Universit{\"a}t Regensburg \\
        E-mail: \email{Semen.Valgushev@physik.uni-regensburg.de}}

\author{M.~Puhr\textsuperscript{*} \\
        Institut f{\"u}r Theoretische Physik, Universit{\"a}t Regensburg \\
        E-mail: \email{Matthias.Puhr@physik.uni-regensburg.de}
        \phantom{\speaker{S.~N.~Valgushev, M.~Puhr}}}

\author{P.~V.~Buividovich\thanks{This work was supported by the S.~Kowalevskaja award from the Alexander von Humboldt foundation.}\\
        Institut f{\"u}r Theoretische Physik, Universit{\"a}t Regensburg \\
        E-mail: \email{Pavel.Buividovich@physik.uni-regensburg.de}}

\abstract{We study the static electric current due to the Chiral Magnetic Effect in samples of Weyl semi-metals with slab geometry, where the magnetic field is parallel to the boundaries of the slab. We use the Wilson-Dirac Hamiltonian as a simplest model of parity-breaking Weyl semimetal with two-band structure. We find that the CME current is strongly localized at the open boundaries of the slab, where the current density in the direction of the magnetic field approaches the conventional value $j = \frac{\mu_5 B}{2 \pi^2}$ at sufficiently small values of the chiral chemical potential $\mu_5$ and magnetic field $B$. On the other hand, very large values of magnetic field tend to suppress the CME response. We observe that the localization width of the current is independent of the slab width and is given by the magnetic length $l_B = 1/\sqrt{B}$ when $\mua \ll \sqrt{B}$. In the opposite regime when $\mua \gg \sqrt{B}$ the localization width is determined solely by $\mua$.}

\FullConference{The 33rd International Symposium on Lattice Field Theory\\
		14 -18 July 2015\\
		Kobe International Conference Center, Kobe, Japan*}

\begin{document}

\section{Introduction}
\label{sec:intro}

 The Chiral Magnetic Effect (CME) \cite{Kharzeev:08:2} and the closely related phenomenon of the negative magnetoresistivity \cite{Nielsen:83:1,Kim:13:1,Kharzeev:14:1,Xiong:15:1,Landsteiner:14:1,Landsteiner:15:1} are nowadays actively studied as the hallmarks of the chiral anomaly in condensed matter systems. Quantitative calculations of these effects require careful regularization both in the infrared and in the ultraviolet, and e.g. in the case of CME different regularizations yield different answers \cite{Burkov:13:1,Vazifeh:13:1}.

 More precisely, the CME can be characterized by the chiral magnetic conductivity $\sigma_{CME}$ which relates the electric current density $\vec{j}$ and the magnetic field $\vec{B}$ and which in general depends on the temporal/spatial modulation of $\vec{B}$:
\begin{eqnarray}
\label{scme_def}
 \vec{j}\lr{w, \vec{k}} = \sigma_{CME}\lr{w, \vec{k}} \, \vec{B}\lr{w, \vec{k}} .
\end{eqnarray}
In this paper, we will consider the static CME response, thus the frequency $w$ will be set to zero in what follows. A continuum calculation of $\sigma_{CME}$ in the static limit using Dirac fermions with Pauli-Villars regularization yields the following result \cite{Buividovich:13:8}:
\begin{eqnarray}
\label{scme_continuum_PV}
 \scme\lr{k} = \frac{1}{\lr{2\pi}^2}\lr{\mu_5 + \frac{\mu_5^2 - k^2/4}{|k|} \log\left|\frac{2\mu_5 - |k|}{2 \mu_5 + |k|} \right|} ,
\end{eqnarray}
where $\mu_5$ is the chiral chemical potential parameterizing the difference between the occupation numbers of right- and left-handed states of Dirac fermions. On the other hand, the continuum calculation without any regularization also yields a finite result which differs from (\ref{scme_continuum_PV}) by a term of the form $\frac{\mu_5}{2 \pi^2}$. The unregularized result implies the existence of the CME current
\begin{eqnarray}
\label{cme_current_naive}
 \vec{j} =  \frac{\mu_5 \, \vec{B}}{2 \pi^2}
\end{eqnarray}
even in the limit $k \rightarrow 0$, that is, for a static magnetic field with infinitely slow spatial variation. On the other hand, the regularized chiral magnetic conductivity in the limit of small momenta $\vec{k}$ tends to zero as $k^2$, thus implying that the CME current should vanish in the infrared \cite{Sadofyev:13:1,Rubakov:10:1}. The naive unregularized value (\ref{cme_current_naive}) is only recovered in the limit of large $k$ \cite{Buividovich:13:8}, albeit with an opposite sign. All lattice calculations of the CME current reproduce the regularized result (\ref{scme_continuum_PV}) up to some lattice artifacts \cite{Buividovich:13:8,Buividovich:15:1,Vazifeh:13:1}.

The vanishing of the static CME current in lattice systems in the infrared immediately rises the question of how such a current can be measured in condensed matter systems subject to strong static and uniform magnetic fields, as in the recent experiments \cite{Kim:13:1,Kharzeev:14:1,Xiong:15:1}. Thus, it is very interesting to consider bounded samples of Weyl semimetal, which is closer to experimental setups. The importance of boundaries was demonstrated in the recent paper \cite{Gorbar:15:1}, where it was shown that in the slab of Weyl semimetal subjected to the external magnetic field perpendicular to the slab boundaries the density of CME current vanishes in contrast to the naive formula (\ref{cme_current_naive}). This is a direct consequence of the finite sample size.

 In these Proceedings we use the simple Wilson-Dirac Hamiltonian with open boundaries to argue that the CME current is still non-vanishing locally even in constant magnetic field parallel to the boundary due to the finite size of the system. More precisely, we find that near the open boundaries of the system the current density is quite close to the naive value (\ref{cme_current_naive}), if the values of $\mu_5$ and $B$ are small. At larger values of $\mu_5$ or $B$  nonlinear effects and lattice artefacts become important, and the CME current saturates. On the other hand, the total current through the whole section of the slab vanishes.

\section{Wilson-Dirac Hamiltonian with open boundaries and the current density operator}
\label{sec:wd_hamiltonian}

 We consider the Wilson-Dirac lattice Hamiltonian on a lattice of slab geometry with open boundaries at $z = 0$ and $z = L_z$. The external magnetic field is directed along the slab in the $x$ direction. The corresponding vector gauge potential is chosen as $A_y = - B z$, so that translational symmetry is intact for both $x$ and $y$ directions. The single-particle Hamiltonian in the coordinate space reads
\begin{eqnarray}
\label{hwdirac_singlepart}
 h_{X, Y}
 =
   \sum\limits_{k=1}^{3} -i \alpha_k \nabla_{k \, X Y}
 +
   \frac{r \, \gamma_0}{2} \Delta_{X Y}
 + \gamma_0 m \delta_{X Y}
 + \gamma_5 \mu_5 \delta_{X Y},
\end{eqnarray}
where $\alpha_k$ with $k = x, y, z$ are the Dirac $\alpha$-matrices, $\gamma_0$ and $\gamma_5$ are the Dirac gamma-matrices in the same representation, capital letters denote a set of spatial coordinates $X = \lrc{x, y, z}$, $\delta_{X Y}$ is the Kronecker symbol for all three coordinates and $\nabla_{k \, X Y}$ and $\Delta_{X Y}$ are the lattice discretizations of the derivative and the Laplacian operators corrected at the open boundaries of the slab:
\begin{eqnarray}
\label{lattice_nabla_delta}
 \nabla_{k \, X Y}
 =
 \frac{1}{2} \lr{\theta\lr{L_z - z - 1} e^{i A_{X,k}} \delta_{X+\hat{k},Y}  - \theta\lr{z} e^{-i A_{X-\hat{k},k}}\delta_{X - \hat{k},Y}} ,
\nonumber \\
 \Delta_{X Y} = \sum\limits_{k=x,y,z}^{3} \lr{2\delta_{X Y} - \theta\lr{L_z - z - 1} e^{i A_{X,k}} \delta_{X+\hat{k},Y} - \theta\lr{z} e^{-i  A_{X-\hat{k},k}}\delta_{X - \hat{k},Y}} ,
\end{eqnarray}
where $\theta\lr{z}$ is the unit step function, which we define as $\theta\lr{z} = 1$ if $z > 0$ and $\theta\lr{z} = 0$ if $z \leq 0$.

 The single-particle current density operators can be obtained in a straightforward way by differentiating the single-particle Hamiltonian (\ref{hwdirac_singlepart})
\begin{eqnarray}
\label{current_density_op}
 j_{k,X,Y}\lr{Z} = \frac{\partial h_{X,Y}}{\partial A_{Z,k}}
 =
 P_+ \theta\lr{z} e^{-i A_{Z,k}} \delta_{Z+\hat{k}, X} \delta_{Z, Y}
 +
 P_- \theta\lr{L_z - z - 1} e^{i A_{Z,k}} \delta_{Z, X} \delta_{Z+\hat{k}, Y},
\end{eqnarray}
where $P_{\pm} = \frac{\alpha_i \pm i \gamma_0}{2}$.

 By virtue of translational invariance in $x$ and $y$ directions the Hamiltonian (\ref{hwdirac_singlepart}) and the current density operator (\ref{current_density_op}) can be partially diagonalized in the basis of plane waves propagating in these directions:
\begin{eqnarray}
\label{hwdirac_singlepart_momentum}
 h_{z z^\prime}\lr{k_x, k_y}
 =
 -i \alpha_3 \nabla_{3 \, z z^\prime} + \frac{r \, \gamma_0}{2} \Delta_{z z^\prime}
 +
 \delta_{z z^\prime} \left[ \alpha_1 \sin\lr{k_x} + \alpha_2 \sin\lr{k_y - B z} \right.
 + \nonumber \\ +
 2 r \, \gamma_0 \sin^2\lr{k_x/2} + 2 r \, \gamma_0  \sin^2\lr{(k_y -B z)/2}
 + \gamma_0 m
 + \left. \gamma_5 \mu_5 \right]
\end{eqnarray}
The partially diagonalized current in the $x$ directions takes a particularly simple form:
\begin{eqnarray}
\label{eq:current_operators}
 j_{x, z z^\prime}\lr{k_x, k_y} = \delta_{z z^\prime}\lr{\alpha_x \cos\lr{k_x} + r \gamma_0 \sin\lr{k_x}}.
\end{eqnarray}

 At zero temperature, the expectation value of the current along the magnetic field is given by the sum of contributions of the eigenstates of the single-particle Hamiltonian (\ref{hwdirac_singlepart_momentum}) with negative energy values (these are the levels in the ``Fermi sea''):
\begin{eqnarray}
\label{eq:current_average}
\vev{j_x}\lr{z}= \sum\limits_i \int\limits^{\pi}_{-\pi}\frac{d k_x}{2\pi}\frac{d k_y}{2\pi} \theta\lr{-\varepsilon_i\lr{k_x, k_y}} \bar{\psi}_{i,z}\lr{k_x, k_y} j_{x, zz}\lr{k_x, k_y} \psi_{i,z}\lr{k_x, k_y} ,
\end{eqnarray}
where the eigenstates $\psi_{i,z}\lr{k_x, k_y}$ and the energy levels $\epsilon_i\lr{k_x, k_y}$ are defined as
\begin{eqnarray}
\label{hwd_eigensys_def}
 \sum\limits_{z^\prime} h_{zz^\prime}\lr{k_x, k_y} \psi_{i,z^\prime}\lr{k_x, k_y} = \varepsilon_i\lr{k_x, k_y} \psi_{i,z}\lr{k_x, k_y} .
\end{eqnarray}

\section{Spatial profiles of the current density}
\label{sec:current_profiles}

 We now calculate the spatial profiles of the CME current (\ref{eq:current_average}) in the direction $z$ perpendicular to the boundaries of the slab. We use the slab width $L_z = 60$. Integration over the longitudinal momenta $k_x$, $k_y$ in (\ref{eq:current_average}) is performed using the \texttt{Cubature} package \cite{Johnson:Cubature}.

\begin{figure*}
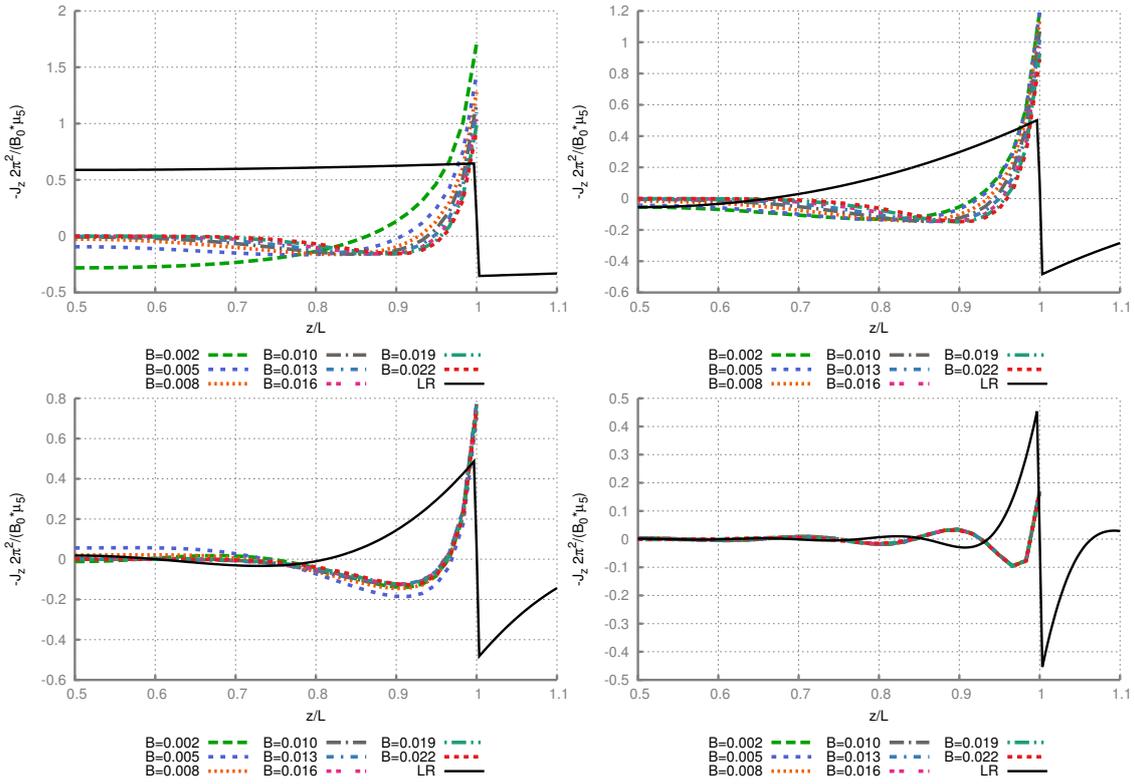

  \centering
  \includegraphics[width=0.49\linewidth]{{{Comp_w_LR_mu0.010}}}\includegraphics[width=0.49\linewidth]{{{Comp_w_LR_mu0.050}}}\\
  \includegraphics[width=0.49\linewidth]{{{Comp_w_LR_mu0.100}}}\includegraphics[width=0.49\linewidth]{{{Comp_w_LR_mu0.300}}}\\
  \caption{Spatial profiles of the current density in the direction of the magnetic field at different values of the chiral chemical potential $\mu_5$ (from left to right and from top to bottom: $\mu_5 = 0.01$, $\mu_5 = 0.05$, $\mu_5 = 0.1$ and $\mu_5 = 0.3$) and the magnetic field $B$. Solid black lines are the result of a linear response theory calculation with $\sigma_{CME}$ given by (\protect\ref{scme_continuum_PV}).}
  \label{fig:spatial_profiles}
\end{figure*}

 On Fig.~\ref{fig:spatial_profiles} we show the spatial profiles of the current density in the direction of the magnetic field at different values of the chiral chemical potential. Since the distribution of the current is completely symmetric with respect to the center of the slab, we show only half of the full profile. One can see that in all cases the current density is concentrated near the boundary of the slab and shows some oscillations of relatively small amplitude in the bulk. The current density directly at the boundary is maximal and is close to the naive value (\ref{cme_current_naive}), but can be both slightly below or slightly above it. Interestingly, the total current through the whole cross-section of the slab seems to vanish within the precision of our integration over the Brillouin zone.

\begin{figure*}
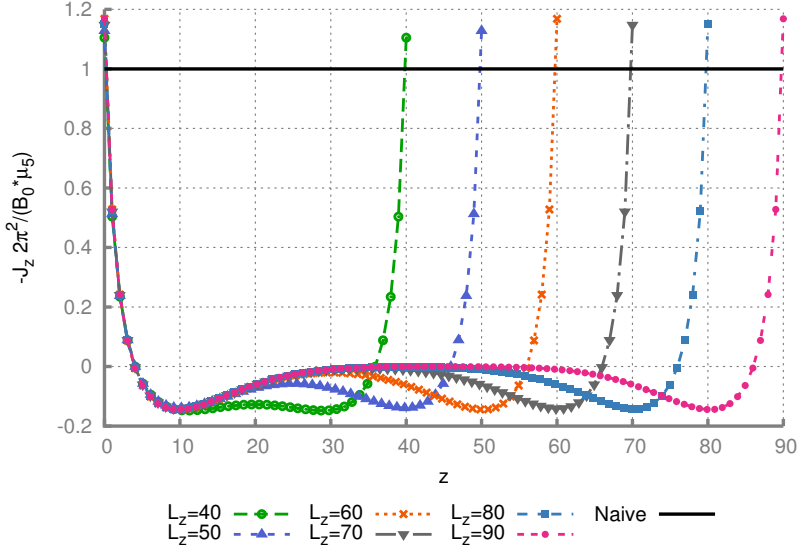

  \centering
  \includegraphics[width=0.69\linewidth]{{{Ldep}}}\\
  \caption{Profile of the current density in the direction of the magnetic field at different values of slab width $L_z$ at fixed values of the magnetic field $B=0.007$ and chiral chemical potential $\mu_5=0.05$. Solid black line represents the naive value of the CME current (\protect\ref{cme_current_naive}). Note that $\sqrt{B} > \mua$.}
  \label{fig:L_dep}
\end{figure*}

 On Fig.~\ref{fig:L_dep} we demonstrate how the profile of the current depends on the slab width $L_z$ at relatively small values of magnetic field and chiral chemical potential.
 It is interesting to note that the localization width and the profile of the current near the boundaries seem to be independent of the slab width. At large slab  widths we observe that the current density deep in the bulk almost vanishes. This suggests that for very large slab widths all non-trivial behavior of the current density is localized near the boundaries and determined solely by the values of the magnetic field $B$ and the chiral chemical potential $\mu_5$.

\begin{figure}
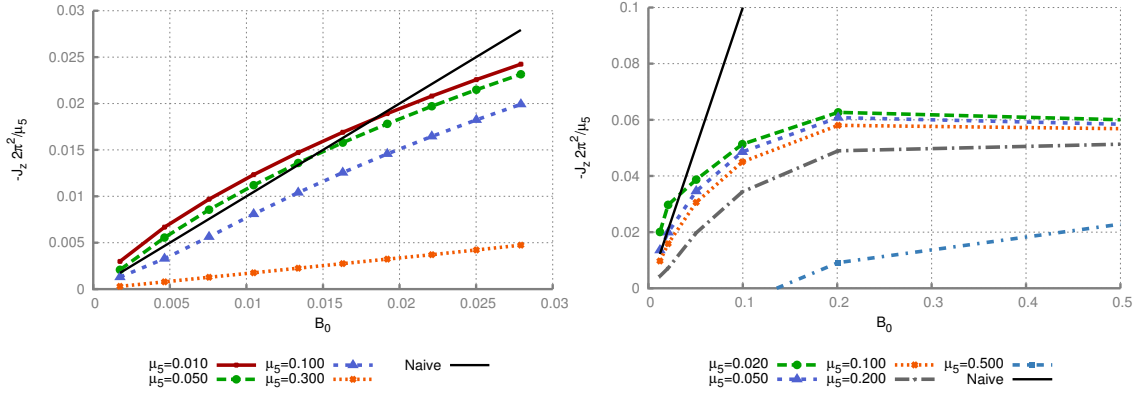

  \centering
  \includegraphics[width=0.49\linewidth]{{{Bdep}}}\includegraphics[width=0.49\linewidth]{{{cBdep}}}\\
  \caption{The dependence of the current density at the boundaries of the slab on the magnetic field strength for several values of the chiral chemical potential $\mu_5$. Solid black line corresponds to the naive result (\protect\ref{cme_current_naive}). The right plot illustrates the saturation of the current at very large values of the magnetic field. }
  \label{fig:Bdep}
\end{figure}

 In order to characterize the dependence of the current density on the magnetic field and the chiral chemical potential $\mu_5$, on Fig.~\ref{fig:Bdep} we plot the current density $j_x$ at the boundaries of the slab (which is also the maximal value of the current density in all the cases which we have considered) as a function of the magnetic field strength at several fixed values of $\mu_5$. At small values of $B$ and $\mu_5$ we see the expected linear scaling with a slope quite close to the naive value (\ref{cme_current_naive}). At larger values of $B$, however, we observe a saturation effect and the current density might even decrease. Assuming a reasonable physical lattice spacing of order of $0.1 \nm$, the values of $B$ at which one sees the saturation would correspond to very large physical magnetic fields of order of hundreds of Tesla, which are certainly out of the reach of modern high magnetic field facilities. Realistic values of $B$ of order of tenth of Tesla are deep in the linear regime, see Fig.~\ref{fig:Bdep} on the left.

 \begin{figure*}
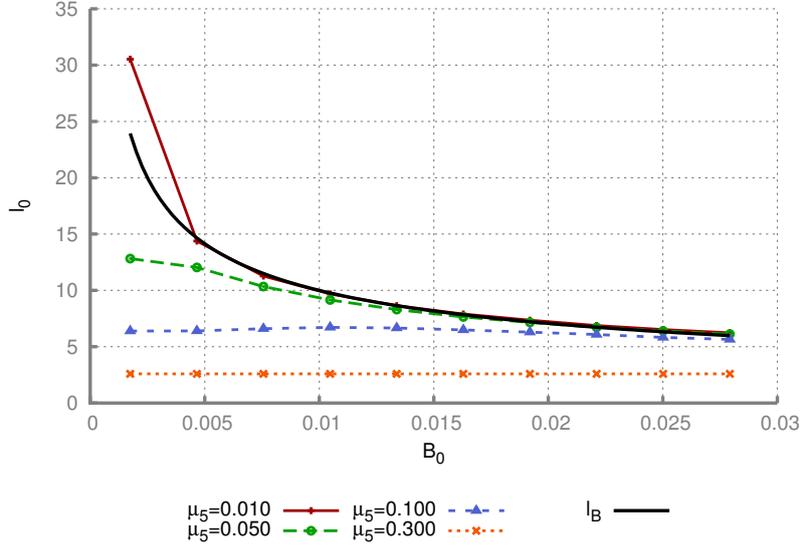

  \centering
  \includegraphics[width=0.69\linewidth]{{{LocWidth}}}\\
  \caption{Dependence of the localization width $l_0$ of the current density on the magnetic field strength at different values of the chiral chemical potential. Solid black line represent the magnetic length $l_B=1/\sqrt{B}$. On this plot $L_z=60$, however it should be emphasized that $l_0$ is completely independent of the slab width when $l_B < L_z/2$. }
  \label{fig:Loc_width}
\end{figure*}

 Next we study the dependence of the localization width of the current density near the boundaries on the strength of the magnetic field $B$ and chiral chemical potential $\mu_5$. We define the localization width $l_0$ as the distance from the boundary to the nearest extremum of the current profile: $\mathrm{d}j_x(z) / \mathrm{d} z \vert_{z=l_0} = 0$. We estimate it numerically for different sets of parameters and present the results on Fig.~\ref{fig:Loc_width}. Interestingly, at small values of the chiral chemical potential $\mu_5 \ll \sqrt{B}$ the width $l_0$ seems to be determined solely by the magnetic field $B$ and agrees with a very good precision with the magnetic length $l_B = 1/ \sqrt{B}$ in the region where $l_B < L_z/2$. However, the chiral chemical potential $\mu_5$ sets a scale roughly $\sqrt{B} \sim \mu_5$ at which the dependence of the width $l_0$ on the magnetic field is saturated and as consequence $l_0$ depends only on $\mu_5$.

  In order to estimate the importance of open boundaries, we also perform a simple comparison with the linearized response (\ref{scme_def}), where the chiral magnetic conductivity is given by the continuum result (\ref{scme_continuum_PV}). To this end we consider a magnetic field that vanishes everywhere except for a finite region of size $L_z$, where it has a nonzero constant value. The fermion mass and the chiral chemical potential are assumed to be constant throughout the whole space. We multiply the Fourier transform of such a step-shaped magnetic field distribution by $\sigma_{CME}\lr{\vec{k}}$ and then transform the resulting current density back into the coordinate space. The result is shown on Fig.~\ref{fig:spatial_profiles} with solid black lines.

\begin{figure*}
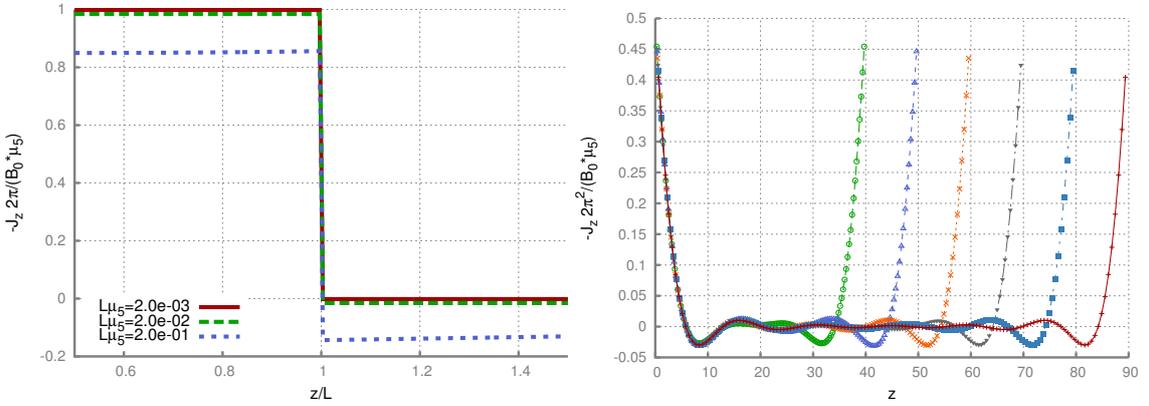

  \centering
  \includegraphics[width=0.49\linewidth]{{{LRcurrent_Lmu_const_small}}}
  \includegraphics[width=0.49\linewidth]{{{LRcurrent_Ldep}}}
  \caption{Profile of the linear response current density. Left: For $L_z\mua \lesssim 1$ the current density inside the slab is constant. Right: If $L_z\mua \gg 1$ the current density is localized near the slab boundary. The localization width $l_0$ does not depend on $L_z$. In the plot $\mua=0.2$ and the current density outside of the slab is not shown.}
  \label{fig:LR_Loc_width}
\end{figure*}

The qualitative behavior of the linear response current density depends on the dimensionless product $L_z \mua$.  For $L_z\mua \lesssim 1$ the current density inside the slab is constant and its magnitude is close to the naive value. In the regime where $L_z \mua \gg 1 $ the current density is localized near the slab boundary. The localization width does not depend on $L_z$ and is  comparable to the one observed in the lattice calculation in the regime with $\mu_5 \gg \sqrt{B}$. Typical results for the linear response current density are shown on Fig.~\ref{fig:LR_Loc_width}.

Let us also note that the current density calculated using the linear response formula only totals to zero when summed over the whole space, not just over the interior of the slab. This is simply because in the linear response calculation we are assuming that outside of the slab we have the same medium, only with vanishing magnetic field.

\section{Discussion}
\label{sec:discussion}
  In this sections we would like to discuss the qualitatively different behavior of the current profile depending on the relation between the chiral chemical potential $\mua$ and the magnetic field $B$.
  First of all, we observe that in the presence of sufficiently strong magnetic field, much greater that $\mua^2$, at least in our simple lattice model nonlinear effects in $B$ become important, whereas in the opposite regime the response of the current is almost linear in the magnetic field.
  Thus it is interesting to estimate which of this two scenarios is realized in our model at typical values of fields from the recent experiments on Dirac and Weyl semimetals.

  To this end we note, that in real experiments the chiral chemical potential is induced dynamically by the anomaly in the background of parallel magnetic and electric fields:
\begin{eqnarray}
\label{anomaly_equation}
\frac{d \rho_5}{dt} = \frac{e^2}{4\pi^2 \hbar^2} E \cdot B - \frac{\rho_5}{\tau_\upsilon},
\end{eqnarray}
where $\rho_5$ is the density of the chiral charge and $\tau_\upsilon$ is the inter-valley scattering time which takes into account mixing between left- and right-handed fermions in the material. We can estimate the induced chiral chemical potential $\mua$ using the following expression for the density of chiral charge at zero temperature and chemical potential \cite{Kharzeev:08:2} :
\begin{eqnarray}
\label{density_of_chiral_charge}
\rho_5 = \frac{\mua^3}{3\pi^2 \hbar^3 \upsilon^3},
\end{eqnarray}
where $\upsilon$ is the Fermi velocity.
Finding a steady state at long times from equation (\ref{anomaly_equation}) and matching the result to the expression (\ref{density_of_chiral_charge}), we find:
\begin{eqnarray}
\label{mua_est}
\frac{\mua}{\upsilon} = 2 \sqrt[3]{\frac{3}{2} \hbar e^2 \tau_\upsilon E \cdot B},
\end{eqnarray}
and for typical values $B \sim 4~T$, $\tau_\upsilon \sim 10^{-14}~ s$, $\upsilon \sim c / 300$ \cite{Kharzeev:14:1} and some reasonable value of electric field $E \sim 10^4~ V\cdot m^{-1}$ we find that
\begin{eqnarray}
\label{E_est}
\frac{\mua}{\upsilon} &\ll & \sqrt{eB\hbar},\nonumber\\
E &\ll & \frac{1}{\tau_\upsilon}\sqrt{\frac{B \hbar}{e}}.
\end{eqnarray}
This suggests that such fields correspond to the nonlinear regime in our lattice model, which is unaccessible from the linear response formula (\ref{scme_def}).

\section{Conclusions}
\label{sec:conclusions}

 We have numerically demonstrated that for a finite-size lattice system of slab geometry the CME current is vanishing \emph{in total}, however, locally the current is still nonzero. It is maximal directly at the boundaries of the slab, where it is quite close to the naive value (\ref{cme_current_naive}). The localization depth of the current seems to be independent of the chiral chemical potential and coincide with the magnetic length $l_B$ when $\mu_5 \ll \sqrt{B}$ and to be mostly determined by the value of the chiral chemical potential $\mu_5$ when $\mu_5 \gg \sqrt{B}$.

 Since the chiral magnetic effect is a crucial ingredient in the negative longitudinal magnetoresistivity phenomenon (NMR) \cite{Kharzeev:14:1}, it would be interesting to calculate the longitudinal conductivity directly using the Hamiltonian (\ref{hwdirac_singlepart}) and understand, whether the current responsible for the NMR is also localized at the boundary of the sample. The discussion presented in Section \ref{sec:discussion} above suggests that in such a calculation the magnetic field cannot be treated as a small perturbation, and one has to calculate the conductivity assuming infinitely small {\it{electric}} field in the presence of strong magnetic field rather that the small {\it{magnetic}} field in the presence of some chiral chemical potential. We leave such a calculation for the future work.


\begin{thebibliography}{10}
\expandafter\ifx\csname url\endcsname\relax
  \def\url#1{{\tt #1}}\fi
\expandafter\ifx\csname urlprefix\endcsname\relax\def\urlprefix{URL }\fi
\providecommand{\eprint}[2][]{\url{#2}}

\bibitem{Kharzeev:08:2}
K.~Fukushima, D.~E. Kharzeev, H.~J. Warringa, {\em The {Chiral Magnetic
  Effect}\/}, Phys.Rev.D {\bf 78} (2008), 074033, \href{http://arxiv.org/abs/0808.3382}{ArXiv:0808.3382}.

\bibitem{Nielsen:83:1}
H.~B. Nielsen, M.~Ninomiya, {\em The {Adler-Bell-Jackiw} anomaly and {Weyl}
  fermions in a crystal\/}, \href{http://dx.doi.org/10.1016/0370-2693(83)91529-0}{Phys.Lett.B {\bf 130} (1983), 389 -- 396}.

\bibitem{Kim:13:1}
H.~J. Kim, K.~S. Kim, J.~F. Wang, M.~Sasaki, N.~Satoh, A.~Ohnishi, M.~Kitaura,
  M.~Yang, L.~Li, {\em Dirac vs. {Weyl} in topological insulators:
  {Adler-Bell-Jackiw} anomaly in transport phenomena\/}, Phys.Rev.Lett. {\bf
  111} (2013), 246603, \href{http://arxiv.org/abs/1307.6990}{ArXiv:1307.6990}.

\bibitem{Kharzeev:14:1}
Q.~Li, D.~E. Kharzeev, C.~Zhang, Y.~Huang, I.~Pletikosic, A.~V. Fedorov, R.~D.
  Zhong, J.~A. Schneeloch, G.~D. Gu, T.~Valla, {\em Observation of the chiral
  magnetic effect in {ZrTe$_5$}\/} (2014), \href{http://arxiv.org/abs/1412.6543}{ArXiv:1412.6543}.

\bibitem{Xiong:15:1}
J.~Xiong, S.~K. Kushwaha, T.~Liang, J.~W. Krizan, W.~Wang, R.~J. Cava, N.~P.
  Ong, {\em Signature of the chiral anomaly in a {Dirac} semimetal: a current
  plume steered by a magnetic field\/} (2015), \href{http://arxiv.org/abs/1503.08179}{ArXiv:1503.08179}.

\bibitem{Landsteiner:14:1}
K.~Landsteiner, Y.~Liu, Y.~Sun, {\em Negative magnetoresistivity in chiral
  fluids and holography\/}, JHEP {\bf 03} (2015), 127, \href{http://arxiv.org/abs/1410.6399}{ArXiv:1410.6399}.

\bibitem{Landsteiner:15:1}
A.~{Jimenez-Alba}, K.~Landsteiner, Y.~Liu, Y.~Sun, {\em Anomalous
  magnetoconductivity and relaxation times in holography\/}, JHEP {\bf 07}
  (2015), 117, \href{http://arxiv.org/abs/1504.06566}{ArXiv:1504.06566}.

\bibitem{Burkov:13:1}
Y.~Chen, S.~Wu, A.~A. Burkov, {\em Axion response in {W}eyl semimetals\/},
  Phys.Rev.B {\bf 88} (2013), 125105, \href{http://arxiv.org/abs/1306.5344}{ArXiv:1306.5344}.

\bibitem{Vazifeh:13:1}
M.~M. Vazifeh, M.~Franz, {\em Electromagnetic response of {Weyl} semimetals\/},
  Phys.Rev.Lett. {\bf 111} (2013), 027201, \href{http://arxiv.org/abs/1303.5784}{ArXiv:1303.5784}.
  
\bibitem{Gorbar:15:1}
E.~V. Gorbar, V.~A. Miransky, I.~A. Shovkovy, P.~O. Sukhachov {\em Chiral separation and magnetic effects in a slab: the role of boundaries\/}
  (2015), \href{http://arxiv.org/abs/1509.06769}{ArXiv:1509.06769}.

\bibitem{Buividovich:13:8}
P.~V. Buividovich, {\em Anomalous transport with overlap fermions\/}, Nucl.
  Phys. A {\bf 925} (2014), 218 -- 253, \href{http://arxiv.org/abs/1312.1843}{ArXiv:1312.1843}.

\bibitem{Sadofyev:13:1}
Z.~V. Khaidukov, V.~P. Kirilin, A.~V. Sadofyev, V.~I. Zakharov, {\em On
  magnetostatics of chiral media\/} (2013), \href{http://arxiv.org/abs/1307.0138}{ArXiv:1307.0138}.

\bibitem{Rubakov:10:1}
V.~A. Rubakov, {\em On chiral magnetic effect and holography\/} (2010),
  \href{http://arxiv.org/abs/1005.1888}{ArXiv:1005.1888}.

\bibitem{Buividovich:15:1}
P.~V. Buividovich, M.~Puhr, S.~N. Valgushev, {\em Chiral magnetic conductivity
  in an interacting lattice model of parity-breaking {Weyl} semimetal\/},
 Phys. Rev. B {\bf 92} (2015), 205122, \href{http://arxiv.org/abs/1505.04582}{ArXiv:1505.04582}.

\bibitem{Johnson:Cubature}
S.~G. Johnson, {\em Cubature (multi-dimensional integration)\/}.
\href{http://ab-initio.mit.edu/wiki/index.php/Cubature}{http://ab-initio.mit.edu/wiki/index.php/Cubature}

\end{thebibliography}
\end{document}